\documentclass[aps,prd,twocolumn,showpacs,preprintnumbers,amsmath,amssymb,floatfix,nofootinbib]{revtex4}
\usepackage{graphicx}
\usepackage{color}
\usepackage{colordvi}

\newcommand{\beqn}{\begin{eqnarray}}
\newcommand{\eeqn}{\end{eqnarray}}
\newcommand{\be}{\begin{equation}}
\newcommand{\ee}{\end{equation}}

\newcommand{\ba}{\begin{array}{c}}
\newcommand{\bat}{\begin{array}{cc}}
\newcommand{\ea}{\end{array}}

\newcommand{\bi}{\begin{itemize}}
\newcommand{\ei}{\end{itemize}}
\newcommand{\chpt}{$\chi$PT}
\newcommand{\rcht}{R$\chi$T}

\newcommand{\cO}{{\cal O}}

\begin{document}

\preprint{FTUV/06-1129}
\preprint{IFIC/06-42}
\preprint{MPP-2006-152}

\title{Vanishing chiral couplings in the large--$N_C$ resonance theory}

\author{Jorge Portol\'es${}^1$}
\author{Ignasi Rosell${}^1$}
\author{Pedro Ruiz-Femen\'\i a${}^2$}

\affiliation{${}^1$ 
IFIC, Universitat de Val\`encia -- CSIC, Apt. Correus 22085, E-46071
Val\`encia, Spain} 
\affiliation{${}^2$  Max-Planck-Institut f\"ur Physik (Werner-Heisenberg Institut), 
F\"ohringer Ring 6, 80805 M\"unchen, Germany 
}

\begin{abstract}
The construction of a resonance theory involving hadrons requires to
implement the information from higher scales into the couplings of the effective Lagrangian.
We consider the large-$N_C$ chiral resonance theory incorporating scalars and pseudoscalars
and we find that, by imposing LO short-distance constraints on form factors of QCD currents 
constructed within this theory, the chiral low-energy constants satisfy resonance saturation
at NLO in the $1/N_C$ expansion.
\end{abstract}



\maketitle


\section{Introduction}

Since the inception of Chiral Perturbation Theory (\chpt) \cite{GL:85} a lot of effort
has been dedicated to the determination of the chiral low-energy constants (LECs),
whether from hadronic observables or through spectral representations
of Green functions that are order parameters of spontaneous chiral symmetry breaking.
However it is well known that the LECs of every effective field theory collect 
information from degrees of freedom that have been integrated out to obtain the 
low-energy Lagrangian. In consequence it has been put forward that chiral LECs would 
receive a contribution from the low-lying resonances that do not appear in \chpt.
This idea has been explored through the construction of a phenomenological
Lagrangian (\rcht) involving one multiplet of vector, axial-vector, scalar and
pseudoscalar resonances \cite{RChTa} and the conclusion that was achieved assesses
the fact that the tree-level integration of the lightest resonance fields 
saturate the phenomenological 
values of ${\cal O}(p^4)$ chiral LECs. An extension of this result up to 
${\cal O}(p^6)$ could be expected \cite{CEEKPP:06}.
\par
The resonance theory can be better understood within the framework of large-$N_C$
QCD \cite{tHO:74} where tree-level interactions between
an infinite spectrum of narrow states implemented in a chiral invariant Lagrangian 
provide the LO ($N_C \rightarrow \infty$)
contribution to Green functions of QCD currents. Thus the idea of matching the 
tree-level functions, evaluated within \rcht ,
with those of QCD in the 
same limit \cite{CEEKPP:06,RChTb,MHA,three} arises naturally and it 
has been shown to broaden widely our knowledge on the construction of the theory by
providing large-$N_C$ estimates of the coupling constants in the Lagrangian that turn
out to be in remarkable agreement with the phenomenology.
\par
The ${\cal O}(p^4)$ couplings in \chpt \ ($L_i$)
and in the theory where the resonances are still active degrees of freedom ($\widetilde{L}_i$)
are related upon integration of the resonance fields:
\begin{equation}\label{saturation}
L_i(\mu) = L_i^R(\mu) + \widetilde{L}_i(\mu) \, ,
\end{equation}
where $L_i^R(\mu)$ is the contribution stemming from the low-energy expansion of
the resonance contributions.
The statement of resonance saturation of the ${\cal O}(p^4)$ \chpt \ couplings
alleges then that $\widetilde{L}_i(\mu)=0$, {\it i.e.} that the values of the LECs are 
generated by the decoupling of the mesonic states which lie above the Goldstone particles.
This assertion immediately raises the question of its validity
for a determined value of $\mu$ or if the result is  accomplished for any value
(\lq \lq extreme" version of resonance saturation~\cite{CP:02}). 
The latter possibility is specially interesting because of its simplicity and naturalness:
the $L_i(\mu)$ couplings are then predicted as a function only 
of the resonance parameters, which can be extracted from the phenomenology or considering
the matching procedure outlined before.
\par
At LO in the $1/N_C$ expansion, 
the asymptotic behaviour of QCD correlators require that 
$\widetilde{L}_i = 0$~\cite{RChTb}, in the \rcht \ formulation where spin-1 mesons are 
described by antisymmetric tensor fields.
In this limit, and considering the large-$N_C$ resonance Lagrangian of Ref.~\cite{CEEKPP:06}, where only contributions from the lightest resonances are taken into account, Eq.~(\ref{saturation}) turns out to be
\begin{align}
&L_1=\frac{G_V^2}{8M_V^2}\,,\quad  
L_2=\frac{G_V^2}{4M_V^2}\,, \quad
L_3=-\frac{3G_V^2}{4M_V^2}\,+\,\frac{c_d^2}{2M_S^2}\, ,\nonumber \\
&L_5=\frac{c_dc_m}{M_S^2}\,,\quad
L_8=\frac{c_m^2}{2M_S^2}\,-\,\frac{d_m^2}{2M_P^2} \,, \quad
L_9=\frac{F_VG_V}{2M_V^2}\,, \nonumber \\
&L_{10}=-\frac{F_V^2}{4M_V^2}\,+\,\frac{F_A^2}{4M_A^2}\,,\qquad
L_4=L_6=L_7=0\,, \label{rescont}
\end{align}
where $F_V$, $F_A$, $G_V$, $c_d$ and $c_m$ are couplings of the R$\chi$T 
Lagrangian \cite{RChTa}.
A $\mu$-dependence in the chiral couplings may appear through quantum corrections. 
Since the $1/N_C$ expansion is equivalent to a semi-classical approximation, there is 
a $1/N_C$ suppression for each loop, therefore
NLO corrections in the large--$N_C$ framework are given 
by one-loop diagrams generated by the \rcht \ Lagrangian.
Studies along this line of research have recently been carried out
 \cite{CP:02,RSP:05,RPP:05,RSP:06,Natxo}. A proper
question that follows is to find out if resonance saturation still holds at
NLO in the $1/N_C$ expansion. It has been pointed out \cite{CP:02} 
the possible connection between resonance saturation and the implementation of 
short-distance constraints in the Lagrangian theory.   
In Ref.~\cite{RPP:05}, using the background field method,
the full one-loop
computation of the $\beta$ function that renormalizes the resonance theory with scalar
and pseudoscalar resonances was performed. 
Indeed one of the main conclusions of that work is that those $\widetilde{L}_i$
related with
the resonance content of the theory do not depend on $\mu$
when short-distance information is used to determine the LO resonance couplings.
\par
In this letter we provide an explanation of the later fact concluding that 
resonance saturation of the $U(3)_L \otimes U(3)_R$
${\cal O}(p^4)$ chiral LECs related with scalar and pseudoscalar resonances 
is satisfied at NLO in the $1/N_C$ expansion
as a consequence of imposing the right high-energy behaviour on 
form factors calculated within the theory.
In particular, we show that
$\widetilde{L}_i(\mu)=0$ for those couplings named as $\widetilde{L}_{4}$, $\widetilde{L}_5$, $\widetilde{L}_8$
in the usual basis of
$SU(3)_L\otimes SU(3)_R$ \chpt~\cite{GL:85}, and for $\widetilde{\alpha}_{18}$~\footnote{$\widetilde{\alpha}_{18}$
is the coupling 
of the operator ${\cal O}_{18}=\langle u_\mu\rangle \langle u^\mu\chi_+\rangle$,
as defined in Ref.~\cite{RPP:05}, which vanishes in the $SU(3)$ case.}. We also 
show the absence of running for the couplings $\widetilde{L}_6$ 
and $\widetilde{L}_7$, though for the latter
we can only conclude that the NLO finite part of the combination 
$\widetilde{L}_6 + \widetilde{L}_7$ must vanish. 

\section{Large-$N_C$ resonance chiral Lagrangian}

The $U(3)_L\otimes U(3)_R$ chiral Lagrangian 
with scalar and pseudoscalar resonance fields 
used in Ref.~\cite{RPP:05} (see also \cite{CEEKPP:06}) 
has, at leading order in $1/N_C$, the structure:
\begin{equation} 
{\cal L}_{\mathrm{R}\chi\mathrm{T}}= {\cal L}^{\chi \mathrm{PT}}_2 + {\cal L}_{\mathrm{kin}}^R
+{\cal L}_2^{R} +{\cal L}_2^{RR}\,,
\label{eq:RchT}
\end{equation}
where $R$ stands for resonance nonets of scalars $S(0^{++})$ or pseudoscalars $P(0^{-+})$.
${\cal L}_2^{\chi \mathrm{PT}}$ is the $U(3)_L\otimes U(3)_R$ ${\cal O}(p^2)$ 
\chpt \ Lagrangian~\cite{GL:85}.
The piece 
${\cal L}_{\mathrm{kin}}^R$ contains the kinetic terms for the
resonance fields and ${\cal L}_2^{R}$ has the generic form
$\langle R\chi^{(2)} \rangle$,
with $\chi^{(2)}$ an ${\cal O}(p^2)$ chiral tensor. 
The second and third terms in Eq.~(\ref{eq:RchT})
yield the most general Lagrangian that can give contributions
to the chiral ${\cal O}(p^4)$ LECs after integrating out 
scalar and pseudoscalar resonances at tree-level~\cite{RChTa,RChTb}. 
Interaction terms among two resonances are included in 
${\cal L}_2^{RR}\sim \langle RR\chi^{(2)} \rangle$,
where $RR=SS,PP,SP$. Upon resonance integration
double-resonance terms
contribute first at ${\cal O}(p^6)$, but
they can be required to satisfy the short-distance behaviour of resonance form
factors~\cite{Natxo}. 
The truncation of the infinite tower of zero-width resonances of 
the large--$N_C$ spectrum to the lowest-lying multiplet, 
as done in~\cite{RPP:05,CEEKPP:06}, 
is not essential in what follows, but can be assumed to ease the discussion.
Likewise, the addition of interaction terms with three
resonances~\cite{CEEKPP:06}
does not change the conclusions of this paper.
\par
Quantum effects can be computed in this large--$N_C$ framework and
yield NLO corrections to tree-level results. 
Dimensional analysis tells us that one-loop diagrams are of 
${\cal O}(p^4,p^2 M_R^2)$, so it
is obvious that additional operators are needed to renormalize 
${\cal L}_{\mathrm{R}\chi\mathrm{T}}$
above~\cite{RPP:05}. Among those, we shall be interested in the counterterms 
from the Goldstone boson Lagrangian of order $p^4$,
${\cal L}^{\mathrm{GB}}_{4} = \sum_i \widetilde{\alpha}_i {\cal O}_i$,
which should be distinguished from the usual \chpt \ Lagrangian expansion, 
${\cal L}^{{\chi\mathrm{PT}}}_{4}$, 
as the couplings of both theories carry information about physics at 
different scales (notice that
we write $\widetilde{\alpha}_i$ as short for all the ${\cal O}(p^4)$ chiral couplings, including
$\widetilde{L}_i$). Resonance saturation at LO translates into the fact
that $\widetilde{\alpha}_i=0$ and then ${\cal L}_4^{\mathrm{GB}}$ vanishes.
At NLO, the absorbed divergences provide a scale dependence in 
the renormalized 
couplings $\widetilde{\alpha}_i(\mu)$, as dictated by the renormalization
group equations~:
\begin{equation}
\mu\frac{d}{d\mu}\widetilde{\alpha}_i \, = \, -\frac{\gamma_i}{16\pi^2}\,.
\label{eq:RGE}
\end{equation}
The $\gamma_i$ are the divergent coefficients of the counterterms
in ${\cal L}^{\mathrm{GB}}_{4}$ and have an explicit dependence
with the couplings of ${\cal L}_{\mathrm{R}\chi\mathrm{T}}$. 
The leading logarithm in the
evolution of the $\widetilde{\alpha}_i$ constant can thus be obtained by plugging
the LO values for the ${\cal L}_{\mathrm{R}\chi\mathrm{T}}$ couplings
inside $\gamma_i$, {\it i.e.} ignoring the $\mu$ dependence on the right-hand-side of 
the RGE equations.
Consequently, a zero value for the divergent part of the $\widetilde{\alpha}_i$ constant automatically
implies that it does not run at one-loop in the large--$N_C$ framework.
\par
By explicit
computation we found \cite{RPP:05} that the divergent part of 
6 out of the 16 ${\cal L}^{\mathrm{GB}}_{4}$ couplings
vanishes after LO predictions
for the constants in ${\cal L}_{\mathrm{R}\chi\mathrm{T}}$ are 
used~\footnote{Actually, there is one more ${\cal L}^{\mathrm{GB}}_{4}$ coupling, $\widetilde{H}_2$, whose
divergent part also vanishes. However, the saturation of the low-energy couplings $H_1$ and $H_2$ by resonances
has no physical significance, as these constants depend on the renormalization scheme used in QCD, and
will not be included in our analysis.}.
The couplings that share this feature 
are the ones accompanying operators with a $\chi_{\pm}$ tensor, 
that are relevant for the renormalization of the 
two-point correlator functions of two scalar or pseudoscalar currents
($\widetilde{L}_6,\,\widetilde{L}_7$ and $\widetilde{L}_8$), and for the scalar form
factors to two Goldstone bosons ($\widetilde{L}_4,\,\widetilde{L}_5$ and $\widetilde{\alpha}_{18}$).
Next we show that
the absence of running for both sets of couplings is
a consequence of enforcing the correct high-energy behaviour in 
the tree-level scalar and pseudoscalar form factors. 

\section{$\langle SS\rangle$ and $\langle PP\rangle$ correlators}

\begin{figure}
\includegraphics[width=6cm]{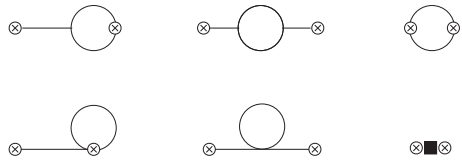}
\caption{Topologies in the one-loop scalar and pseudoscalar correlators. The lines
can represent both Goldstone and resonance fields.\label{fig:correlators}}
\end{figure}

Let us consider the two-point functions built from two scalar $(SS)$ or two
pseudoscalar $(PP)$ currents. Their tree-level expressions are given by
one-particle exchanges,
so they are booked as ${\cal O}(q^{-2})$ at large energies, being $q$ the momentum flowing 
into the current vertex. The 
topologies that arise at one-loop from the Lagrangian in Eq.~(\ref{eq:RchT}) thus yield the ${\cal O}(q^0)$
contributions that are
shown in Fig.~\ref{fig:correlators}. 
The ${\cal L}_4^{\mathrm{GB}}$ operators $\langle \chi_{+}^2\rangle,\,\langle \chi_{+}\rangle^2$ and
$\langle \chi_{-}^2\rangle,\,\langle \chi_{-}\rangle^2$ 
also contribute through local counterterm diagrams (see Fig.~\ref{fig:correlators}), and
their divergent parts 
are fixed uniquely by the renormalization of the $SS$ and $PP$ correlators, respectively.
Other diagrams with
counterterms connected to the external currents with one or two propagators
may also be required in order to absorb all the divergences
from the one-loop graphs. Among the latter,
note that the divergences arising from tadpoles do not 
play any role in the determination of the local counterterms.
\par
The relevant topologies involve loops with two propagators.
After reduction to scalar
integrals, all terms are proportional to the scalar two- and 
one-point functions 
$B_0(q^2,M^2,M^{\prime 2})$ and $A_0(M^2)$ \cite{Passarino:1978jh},
with $M,\,M^\prime$ any of
the masses inside the loops. The divergences
that have to be canceled by the local counterterms of
${\cal L}^{\mathrm{GB}}_{4}$ are the ones proportional to ${\cal O}(q^0)$.
Due to the fact that the $1/\epsilon$ terms from the one-point scalar function are 
proportional to a mass squared, it is easy to convince oneself that
the ${\cal O}(q^0)$ divergences in the $SS$ and $PP$ correlators come
solely from the two-point functions $B_0$.
\par
The spectral functions of the scalar and pseudoscalar
correlators are generated from the discontinuities of the two-point functions.
Using the optical theorem, the spectral function
can be written as a sum over the form factors of all 
absorptive contributions:
\begin{equation}
\mbox{Im} \Pi(q^2) = \sum_n \xi_n(q^2)\left|{\cal F}_n(q^2) \right|^2\,.
\label{eq:spectral}
\end{equation}
At one-loop, any of the possible absorptive contributions, $n$, comes from
the two-particle cuts in the diagrams of Fig.~\ref{fig:correlators}.
If we stick to the particle content in ${\cal L}_{\mathrm{R}\chi\mathrm{T}}$,
the terms in the sum correspond to $n=\phi\phi,\,R\phi,\,RR$, where $\phi$ denotes
a Goldstone boson and $R=S,\,P$ is a resonance field.
The one-loop spectral function is thus entirely determined by 
the tree-level scalar and pseudoscalar form factors to these two-particle states.
It is a commonly accepted statement that the individual
form factors of QCD currents
should vanish at infinite momentum transfer~\cite{Lepage:1980fj}. In  
\rcht \  the appropriate high-energy behaviour is guaranteed by the 
well-known relations among the resonance couplings at LO in the large--$N_C$ limit.
Since the kinematic factors $\xi_n(q^2)$ behave as ${\cal O}(1)$ in the $q^2\to\infty$
limit for the
allowed two-particle cuts from ${\cal L}_{\mathrm{R}\chi\mathrm{T}}$,
the short-distance behaviour of the form factor leads
immediately to a vanishing ${\cal O}(q^{0})$ term for the spectral functions. 
As the ${\cal O}(q^{0})$ absorptive and divergent parts of the correlators
come together in the $B_0$'s, it follows that they are affected by the same suppression.
We therefore reach the conclusion that the divergent
${\cal O}(q^{0})$ piece of the $SS$ and $PP$ correlators, responsible for the
running of $\widetilde{L}_6,\,\widetilde{L}_7$ and $\widetilde{L}_8$, must vanish if the tree-level
scalar and pseudoscalar  form factors computed from the theory behave
as $1/q^2$ at large $q^2$. In more physical terms, imposing the right
short-distance properties at the Lagrangian level produces ultraviolet
finite results for the ${\cal O}(q^{0})$ correlators so that the
renormalization of the local terms is not needed.
\par
The whole argument above can be simplified as follows. If we expand the
correlator in $q^2$, we realize that the ${\cal O}(q^{0})$ terms from the different
one-loop diagrams are either zero or proportional to a unique function, 
$B_0(q^2,0,0)$, so that
\begin{eqnarray}
\Pi( q^2\to\infty)&=& \lambda \,B_0(q^2,0,0)+\cO\left(q^{-2}\right)\,,
\label{eq:Pi}
\end{eqnarray}
with $\lambda$ a combination of resonance parameters.
When we impose relations among the couplings so 
that the imaginary part of the correlators vanishes, we are indeed  
setting $\lambda =0$. This cancels 
out the whole ${\cal O}(q^{0})$ term, including the $1/\epsilon$ and
the finite parts. The saturation of the couplings at NLO in the large--$N_C$ 
counting is thus complete: the running is zero and a local NLO finite piece
from the $\widetilde{L}_6+\widetilde{L}_7$ and $\widetilde{L}_8$ 
couplings is not allowed because of
its wrong high-energy behaviour, since for massless quarks the
correlator $SS-PP$ vanishes as $1/q^4$~\cite{SVZ:79}.
The absence
of a NLO piece from $\widetilde{L}_8$ in ${\cal L}_{\mathrm{R}\chi\mathrm{T}}$
is consistent with a recent determination of the
\chpt \ low-energy coupling $L_8(\mu)$~\cite{RSP:06}.
We would like to point out that the result in Eq.~(\ref{eq:Pi}) is not modified
if an arbitrary number of resonance multiplets is considered, provided their interactions follow
the structure given by the Lagrangian in Eq.~(\ref{eq:RchT}).

\section{Scalar form factor}

\begin{figure}
\includegraphics[width=7.5cm]{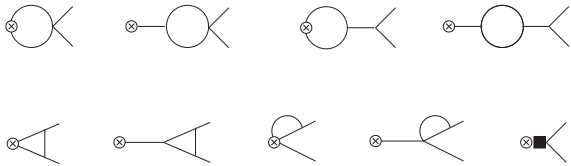}
\caption{Topologies in the one-loop scalar form factor to two Goldstone bosons. Tadpole 
diagrams have not been drawn.\label{fig:scalarformfac}}
\end{figure}

Similarly, counterterms of the operators $\widetilde{L}_4,\,\widetilde{L}_5$ and $\widetilde{\alpha}_{18}$
can be determined by the renormalization of the scalar form factor of two Goldstone 
bosons.
At one-loop the form factor behaves as $q^2$ at large energies, and the allowed
topologies are shown in Fig.~\ref{fig:scalarformfac}. 
The last diagram 
represents the local counterterms of $\widetilde{L}_4,\,\widetilde{L}_5$ and $\widetilde{\alpha}_{18}$
that absorb the ${\cal O}(q^2)$ divergences.
\par
We shall prove first that the ${\cal O}(q^2)$ term of the one-loop
calculation is only proportional to $q^2\,B_0(q^2,0,0)$. For the bubble topologies
(diagrams in the first line of Fig.~\ref{fig:scalarformfac}) this is 
inferred from the discussion above.
A new feature arises from the three propagator integrals.
After the reduction of the one-loop diagrams with three propagators is done, 
the leading term in the $q^2\to\infty$ limit can only be 
proportional to $q^2\,B_0(q^2,0,0)$ or, a priori, to $q^4\,C_0(q^2,0,0,0)$, based on
pure dimensional grounds and on the fact that the 
scalar three-point function, $C_0$, behaves as $1/q^2$.
However it is easy to show that no terms proportional to $q^4\,C_0$
can arise from the triangle loops. Choose the routing of the loop momentum $k$ such that
it is assigned to the vertical line in the triangle. 
The ${\cal O}(p^2)$ vertices connected to the outgoing Goldstone bosons, with momenta $p_1$ and $p_2$,
can thus yield $p_1 \cdot k,\,p_2 \cdot k,$ or $k^2,\,p_1^2,\,p_2^2$ factors.
Take, for example, the upper outgoing line
to be $p_1$, and write the upper vertex factors as 
$p_1 \cdot k=1/2[(k+p_1)^2-k^2-p_1^2]$ and $k^2=(k^2-M^2) +M^2$, with $M$ the mass of the particle
in the vertical propagator. These factors then give either one square mass
term 
multiplying
the three-propagator integral, or has the structure of one of the propagators 
joining at the vertex. 
In the latter case one gets two-propagator integrals that yield $B_0$ or $A_0$ functions. 
In particular, only the two-point function which arises when the vertical propagator is 
canceled out
can pick an additional $q^2$ from the other vertex and
yield a $q^2 B_0$. On the
other hand, a
scalar three-point function only survives if we pick the mass squared term
from each vertex. We thus conclude that $C_0$ enters the result with a $M^4$ factor in 
front. Possible one-point functions $A_0$ do not contribute 
to the leading order in $q^2$ either,
since they are proportional to a square mass. The same is true for the last two one-loop 
diagrams in Fig.~\ref{fig:scalarformfac}. Consequently, the behaviour of the 
scalar form factor of two Goldstone bosons at large energies reads
\begin{eqnarray}
\mathcal{F}(q^2\to\infty)&=& \lambda' \,q^2\,B_0(q^2,0,0) + \cO\left(q^0\right)\,,\label{kp}
\end{eqnarray}
being $\lambda'$ a combination of resonance parameters.
\par
Now consider the absorptive contributions of the one-loop diagrams.
According to the discussion above, only the two-particle cuts in the $s$-channel 
contribute to the ${\cal O}(q^2)$ imaginary part of the scalar form factor, proportional to $q^2 B_0$.
The optical theorem states that the one-loop form factor into two Goldstone bosons is 
given
by the sum of the tree-level form factors to all possible intermediate states 
times the conjugate tree-level scattering amplitude of the intermediate state to two 
Goldstone (which is of ${\cal O}(q^2)$ in the $q^2\rightarrow\infty$ limit),
\begin{eqnarray}
\mbox{Im} \mathcal{F}(q^2) & = &\sum_n \xi_n(q^2)\,{\cal F}_n(q^2)\, A_{\mathrm{scatt}}^*(q^2)\,.
\label{eq:spectralbis}
\end{eqnarray} 
If the tree-level form factors ${\cal F}_n$ obey the $1/q^2$ suppression, 
we conclude that the ${\cal O}(q^2)$ term of $\mbox{Im} \mathcal{F}$ must vanish, and
therefore $\lambda'=0$ in Eq.~(\ref{kp}).
Consequently there is neither an ${\cal O}(q^2)$ divergence to be absorbed by the local 
counterterms of $\widetilde{L}_4$, $\widetilde{L}_5$ and $\widetilde{\alpha}_{18}$, nor any ${\cal O}(q^2)$ finite piece 
coming 
from the loops. A possible NLO finite piece from the $\widetilde{L}_4$, $\widetilde{L}_5$ and $\widetilde{\alpha}_{18}$ 
operators
cannot thus be canceled by possible loop contributions, and it is not allowed
if the scalar form factor to two Goldstone bosons has to obey the $1/q^2$ behaviour at 
NLO in the
large--$N_C$ counting.

\section{Other ${\cal L}_4^{\mathrm{GB}}$ couplings}

It is tempting to apply the preceding discussion to study the renormalization
of the vector-vector and axial-vector--axial-vector correlators, and to the 
vector form factor into
two Goldstone bosons, since they are the key objects to determine the 
divergent piece of the couplings $\widetilde{L}_9$ and $\widetilde{L}_{10}$. This requires 
the 
introduction of vector and axial-vector meson fields in the large--$N_C$ Lagrangian,
which can be done systematically~\cite{RChTa,CEEKPP:06}. 
A problem, however, arises from the fact that
the spin-1 field propagator behaves as ${\cal O}(q^0)$ at large $q^2$
and breaks the $q^2$-counting
advocated before for scalar and pseudoscalar resonances. 
This fact can produce one-loop terms that are higher than ${\cal O}(q^2)$ enhanced
with respect to the tree-level ones when spin-1 resonances flow inside the
loops (see {\it e.g.} the one-loop vector form factor computation
in Ref.~\cite{RSP:05}). The proof given above applies only to 
the leading order divergence for large $q^2$ associated to 
each intermediate state. Thus from the loops which involve cuts with spin-1 resonances,
we can only conclude that their contributions to
the divergent part of
certain ${\cal L}^{\mathrm{GB}}_{6,8}$ couplings vanish if the 
corresponding tree-level 
form factors have the right short-distance suppression.
The cancellation of the subleading divergent term, relevant for the 
renormalization of the ${\cal L}^{\mathrm{GB}}_{4}$ operators, is more subtle for the
loops which involve cuts with spin-one resonances, and very likely
requires a detailed study of the allowed vertex structures~\cite{IPJ}. For the rest of
${\cal L}^{\mathrm{GB}}_{4}$ couplings, namely $\widetilde{L}_1$, $\widetilde{L}_2$, $\widetilde{L}_3$ and 
$\alpha_3$, $\alpha_4$, $\alpha_{17}$~\cite{RPP:05}, that are relevant for the 
renormalization of the elastic Goldstone boson scattering amplitude at one-loop,
we can expect that the analysis
of the high-energy behaviour of the tree-level scattering amplitude of
Goldstone bosons to the possible intermediate states could yield
constraints on the running of these couplings, 
but at the moment this is just a desirable conjecture.
\par
In conclusion we have established that those $U(3)_L \otimes U(3)_R$ chiral
LECs of the large-$N_C$ resonance theory related with scalar and pseudoscalar
resonances do not run at NLO when the theory is properly devised, {\it i.e.}
once the right high-energy behaviour of form factors 
has been implemented by tuning the couplings of the resonance theory at LO.
In between we also conclude that any NLO finite contribution to 
$\widetilde{L}_{4,5,8}$ and $\widetilde{L}_6 + \widetilde{L}_7$ should also
vanish. This outcome
together with the LO result ($\widetilde{L}_i = 0$)
confirms the statement of resonance saturation of chiral LECs up to NLO. 

\acknowledgments

We wish to thank G.~Amor\'os, G.~Ecker, S.~Peris, A.~Pich and J.J.~Sanz-Cillero
for their useful comments on the manuscript.
I.R. is supported by a FPU contract (MEC). This work has been supported in part by the EU
MRTN-CT-2006-035482 (FLAVIAnet), by MEC (Spain) under grant
FPA2004-00996 and by Generalitat Valenciana under grants ACOMP06/098 and GV05/015.
\vspace*{0.1cm}


\end{document}